# Lasing oscillation condition and group delay control in gain-assisted plasmon-induced transparency

Zi-Lan Deng,[1,2] Jian-Wen Dong,[1,b)] He-Zhou Wang,[1]
S. H. Cheng,[3] and Jensen Li [2, a)]

[1] State Key Laboratory of Optoelectronic Materials and Technologies, Sun Yat-Sen (Zhongshan) University, Guangzhou 510275, China

[2] Department of Physics and Materials Science, City University of Hong Kong, Tat Chee Avenue, Kowloon Tong, Hong Kong

[3] Department of Biology and Chemistry, City University of Hong Kong, Tat Chee Avenue, Kowloon Tong, Hong Kong

A gain-assisted plasmonic waveguide with two detuned resonators is investigated in the plasmon-induced transparency window. Phase map is employed to study power transmittance and group delay for varying gain coefficients and frequency detunings of the two resonators. The gain coefficient for lasing oscillation condition is analytically shown to vary quadratically with the frequency detuning. In the amplification regime below the lasing threshold, the spectrum implies not only large group delay, but also high transmittance and narrow linewidth. This is in contrast to those in the loss-compensation regime and the passive case in which there always exists a trade-off between the linewidth and the peak transmittance.

a) Electronic mail: jensen.li@cityu.edu.hk
b) Electronic mail: dongjwen@mail.sysu.edu.cn

The concept of electromagnetically induced transparency originates from atomic physics in which a coherently prepared optical medium becomes transparent in a narrow frequency band in the presence of a pumping laser field.[1] The steep frequency dispersion associated with the induced transparency leads to the slowing down or even completely stopping of light pulses in the medium.[1, 2] Induced transparency can be interpreted from two alternative physical pictures. It comes from the destructive interferences between different excitation pathways of the bare states or equivalently comes from the interferences of the two decay channels of detuned resonances.[3] These two pictures enable to find analogies in many classical optical systems such as coupled micro-resonators,[4] photonic crystal cavities[5, 6] and recently plasmonics[7-11] and metamaterials[12-14]. There are a broad range of applications such as slow light and biosensing. Since surface plasmon polariton (SPP) can confine optical waves in subwavelength scale,[15] plasmonic analogy of induced transparency, i.e. plasmon-induced transparency (PIT), has the device size in nanoscale, superior to the coupled micro-resonators and photonic crystal cavities in micrometer scale. However, intrinsic absorption loss of metal limits the performance of the PIT transmission peaks, in particular to those with high quality factor. A representative example is in the system with detuned resonators. The PIT peak's linewidth will be narrow by decreasing the frequency detuning of two resonators,[8, 10] but its transmittance will be low and eventually disappear when the detuning is small enough. Similar trade-off can be found in another PIT system derived from bright-dark modes interaction.[7, 13] As a result, a linewidth as narrow as that in dielectric system is seldom observed in plasmonic system, and thereby it limits the performance in the field of slow light or biosensing. To mitigate the loss problem, one of the most promising strategies is to incorporate gain materials in close proximity to the metal surface.[16, 17] But there are seldom available models to understand the fact how the loss will be compensated and how it will amplify in the gain-assisted PIT system. Our motivation is to find the loss-compensation and amplification behaviors for different gain coefficients and frequency detuning in gain-assisted PIT system, in order to well control the surface plasmon propagation, and to investigate the condition of plasmonic lasing for the detuned resonators.

In this Letter, we consider a gain-assisted plasmonic waveguide side-coupled with a pair of slightly detuned resonators. We engineer power transmission and group delay by employing a phase map on the gain coefficient and the frequency detuning. It is found that the gain coefficient at the lasing condition in the map can be well-defined by a quadratic function, which is consistent with the analytical derivation in coupled mode theory incorporated with gain coefficients. In the amplification regime below the lasing threshold, both transmittance and linewidth will increase with the decrease of the detuning, in contrast to the trade-off between peak sharpness and peak transmittance in the loss-compensation regime and the passive case. These findings are promising for the fact that the SPP wave would highly transmit along the slow-light waveguide with large group delay, facilitating on-chip optical signal processing with plasmonic elements.

Figure 1(a) shows the schematic of the structure that exhibits the PIT behavior for SPP waves propagating along the x direction. The metal substrate and an adjacent dielectric layer with the thickness of $d$ form a SPP waveguide for transverse magnetic (TM) polarization. Two metallic stripes with same height (w) but slightly different lengths ($L_1$ and $L_2$), placing on top of the dielectric layer, serve as two plasmonic resonators with slightly detuned resonant frequencies. The center-to-center distance between two stripes (L) is tuned to satisfy PIT resonance condition (L ~ 643 nm to have a round trip phase of $2\pi$). The metal is considered as silver whose permittivity is described by Drude model $\varepsilon_{Ag}(\omega) = \varepsilon_\infty - \omega_p^2/\omega(\omega+i\gamma)$, with $\varepsilon_\infty = 3.7$, $\omega_p = 9.08$ eV, and $\gamma = 0.018$ eV. The refractive index of the dielectric layer is taken as 3.53 and optical gain is introduced into part of the dielectric layer (orange region in Fig. 1) to compensate radiation and dissipation losses. In the gain region, we assume the complex refractive index with real part $n_d' = 3.53$, while its imaginary part becomes negative when gain is introduced (through pumping) and is modeled by $n_d'' = -\lambda\alpha(\lambda)/2\pi$.[14] We assume the gain coefficient $\alpha(\lambda)=\exp(-(\lambda-\lambda_0)^2/\ln2(FWHM)^2)$ has a Gaussian profile versus wavelength with the maximum $\alpha_0$ at $\lambda_0 = 1300$ nm and FWHM = 300 nm. These parameters are close to GaAs-based semiconductors doped with quantum dots or quantum wells. As the maximum values of the $E_x$ fields of the SPP guided mode is on the dielectric-air interface, two stripes have a chance to side-couple to the SPP waveguide with high efficiency so that the PIT behavior occurs. A representative PIT transmission spectrum and the corresponding field distributions near the stripes, are illustrated in Fig. 1(b). The field enhancement near both stripes is significantly strengthened at the PIT peak, ten times to the field enhancement in the individual stripe at the left/right transmission dip. We utilize finite element method implemented by COMSOL package to simulate the propagating characteristics of the SPP mode. Complex transmission and reflection coefficients $t$ and $r$ are determined when SPP propagates through plasmonic resonators.

In order to investigate the transmission properties when introducing gain material, we first compute the power transmission for various combinations of the difference in lengths of two stripes ($\Delta L=L_2-L_1$) and the gain coefficient ($\alpha_0$), as shown in Fig. 2(a). We see that the bright color (highlighted by the dash line) forms a remarkable trajectory in the phase map. It means that the transmission is ultra high when the combination of ($\Delta L, \alpha_0$) at the trajectory is chosen. This trajectory is corresponding to the Fabry-Perot oscillation condition for lasing, and can be used for the guideline of determining the lasing threshold (of the gain coefficient) for a particular value of the difference in lengths of two stripes. Actually, one can fit the blue dash line in Fig. 2(a) using quadratic function with the form of

$$\alpha_0 = A\,\Delta L^2 + B \quad (1)$$

where $\Delta L=L_1-L_2$, and $A=20.8$, $B=4.0\mathrm{e}4$ are fitting parameters. As $\alpha_0$ is proportional to the imaginary part of the propagation constant of the SPP mode, and $\Delta L$ is proportional to the frequency detuning. Eq. (1) indicates that smaller detuning will lead to smaller threshold for lasing. For example, one can choose zero frequency detuning (i.e. same lengths of both stripes) to obtain the lower lasing threshold.

The lasing condition for the PIT transmission map shown in Fig. 2(a) can be well described using a coupled mode theory. We assume the transmission/reflection spectra of the single resonator coupled to the SPP waveguide have the simple Lorentz line shape given by

$$t_{1,2}(\omega)=\frac{i(\omega-\omega_{1,2})-\gamma_{1,2}}{i(\omega-\omega_{1,2})-(\gamma_{c1,2}+\gamma_{1,2})},\ r_{1,2}(\omega)=\frac{\gamma_{c1,2}}{i(\omega-\omega_{1,2})-(\gamma_{c1,2}+\gamma_{1,2})} \quad (2)$$

where $\omega_{1,2}$ is the resonant frequency of each individual resonator, and is determined by the length of stripe $L_{1,2}$. $\gamma_{1,2}$ is the decaying rate to outer environment. $\gamma_{c1,2}$ is the coupling rate between the resonators and the waveguide. When two resonators are side-coupled to the SPP waveguide, the transmission spectra can be written by utilizing the Fabry-Perot (FP) model,

$$t_{FP}(\omega) = \frac{t_1(\omega)t_2(\omega)}{1-|r_1(\omega)||r_2(\omega)|\exp(i\phi_{rt}-2\beta''L)} \qquad (3)$$

where $\phi_{rt} = arg(r_1) + arg(r_2) + 2Re(\beta(\omega))L$ is the accumulated round trip phase change. $\beta(\omega)$ is the propagation constant for the propagating SPP mode. $\phi_{rt}$ equals to $2n\pi$ at the PIT peak. When gain is introduced, $\beta(\omega)$ will have a negative imaginary part (i.e. $\beta=\beta'+i\beta''$, $\beta''<0$), as $\beta''$ is related to gain coefficient $\alpha_0$ of the gain material. Without loss of generality, we set $\gamma_1=\gamma_2=\gamma$, $\gamma_{c1}=\gamma_{c2}=\gamma_c$, $\omega_1=\omega_0-\Delta\omega/2$, $\omega_2=\omega_0+\Delta\omega/2$ for simplicity (we can extract $\gamma_1$ and $\gamma_2$ from the numerical simulations of the transmittance and reflectance of single unit but they are similar since $|\Delta L|<<L_{1,2}$). The term $\Delta\omega$ is related to the length difference $\Delta L$ [the x-axis in Fig. 2(a)]. After some derivations, we have a simple expression for the transmittance at the PIT peak,

$$T_{max} = |t_{FP}(\omega=\omega_0)| = \left(\frac{e^{-\beta''L}(\Delta\omega^2+4\gamma^2)}{\Delta\omega^2 - 4e^{-2\beta''L}\gamma_c^2 + 4(\gamma+\gamma_c)^2}\right)^2 \qquad (4)$$

Using Eq. (4), we plot another phase map for various combinations of ($\Delta\omega$, $-2\beta''L$), which is tightly related to ($\Delta L, \alpha_0$). One can find that the trajectory of the lasing condition still exists and is similar to COMSOL simulation in Fig. 2(a). The trajectory is the case when the denominator of Eq. (4) vanishes, yielding

$$-2\beta''L = \Delta\omega^2/4\gamma_c^2 + (\gamma/\gamma_c+1)^2 \qquad (5)$$

where $e^{-2\beta''L} \approx -2\beta''L$ is applied as $-2\beta''L \ll 1$. Eq. (5) is the quadratic function in the same form of Eq. (1). When Eq. (5) is satisfied, the transmittance diverges and the amplification reaches the lasing threshold. It indicates that the transmittance in COMSOL simulation [points on the blue dashed curve in Fig. 2(a)] will go to infinity in principle. However, we emphasize that the transmittance will not diverge in real physics system as the gain materials always have saturation and depletion when lasing happens. The lasing condition in the transmission map gives guidance for us to determine the lasing threshold. The actual lasing dynamics in a real physical system can be studied by time domain methods.[18, 19] We also note that the transmittance will decrease with the increase of gain coefficient inside the quadratic curves in Figs. 2(a) and 2(b), which is consistent with previous results in literature using frequency domain method.[20]

We note that the transmission peak near the lasing threshold can survive in the gain system even if the frequency detuning becomes zero. This is in contrast to the fact that the peak will completely disappear in the absence of gain, see Fig. 3(c) for example. But it can be understood by Eq. (4). As the resonators have decay channels to outer environment (by radiative loss and dissipative loss) in plasmonic system ($\gamma \neq 0$), it leads that the numerator in Eq. (4) is always finite even if $\Delta\omega=0$, and $T_{max}$ could still diverge when the denominator in Eq. (4) went to zero. As a result, the amplification and lasing may still occur when the resonators are identical.

Below the lasing threshold, the behavior of the PIT transmission is rich in different region of the map. To illustrate more clearly, we plot the transmission spectra near the PIT window of the waveguide at the given gain coefficients and length differences. The results are shown in

Fig. 3. The red, blue, and black curves are corresponding to different length detuning, $\varDelta L$=4, 2, and 0 nm, respectively. Figure 3(a) shows the PIT transmission spectra at $\alpha_0$=1156 cm$^{-1}$, corresponding to the amplification regime in Fig. 2. The amplification regime is defined by the region bounded by the "$T_{max}$=1" lines (green dash-dot in Fig. 2) and the lasing condition (blue dashed in Fig. 2), where "$T_{max}$=1" means unity peak transmittance. The transmittance increases with the decreasing of the frequency detuning, in contrast to the case in the passive system. In passive system, it is well-known that the PIT peaks will disappear smoothly as the detuning goes to zero. In loss-compensation regime (below the green dash-dot line), e.g. $\alpha_0$=1100 cm$^{-1}$ in Fig. 3(b), the PIT peaks will not vanish but still decrease as the detuning goes to zero. Other cases in loss-compensation regime are plotted in Fig. 2(a). One can see that there is a dark area below the "$T_{max}$=1" line. This is similar to the results calculated by couple mode theory in Fig. 2(b). We notice that such dark area, as well as the "$T_{max}$=1" line, is asymmetric with respect to the y axis in Fig. 2(a), while they are symmetric in Fig. 2(b). The distinction between simulation and theoretical model is because the transmission spectrum of single plasmonic resonator is of asymmetry Fano-type, instead of the exact Lorentz assumption [Eq. (1)] in the model.

Next, we study the group delay properties at the frequency of the PIT peaks by utilizing the couple model theory. The group delay is defined by the dArg(t)/dω, where d(.) is the derivative operation, and Arg(t) is the phase of transmission coefficient. This is followed by the definition of Ref. 5. We calculated the group delay map in the plane of ($\Delta\omega$, $-2\beta''L$), as illustrated in Fig. 4(a). The group delay increases below the lasing threshold by either introducing enough gain or reducing detuning. For a given gain material, it means that the group delay and the transmittance of the SPP wave can simultaneously increase even in zero detuning. Consequently, the SPP wave would highly transmit along the slow-light waveguide with large group delay, compared to the bare structure without metallic stripes. On the other hand, it also indicates that one may change the gain coefficient (through pumping power) to tune the delay time and potentially to realize tunable delay line for SPP wave. Figure 4(b) plots the relationship between the group delay and the term $-2\beta''L$ related to the gain coefficient. It is clear that all color curves rise obviously. To demonstrate the above concepts, we performed numerical simulations to calculate the group delay of the coupled detuned resonators shown in Fig. 1. We plot the case on zero detuning in Figs. 4(c) and 4(d). The group delay time reaches 2ps for $\alpha_0=1100\text{cm}^{-1}$ and 70ps for $\alpha_0=1156\text{cm}^{-1}$, which are much larger than that of bare structure (approximate to 0.02ps). We also note that the curves near and beyond the lasing condition (e.g. dash curves in Fig. 4(b)) are meaningless because the frequency domain results may not imply the real physics due to the gain saturation and depletion is not taken into account.

In summary, we investigate a sub-wavelength plasmonic waveguide with slightly detuned resonators in the window of plasmon induced transparency. Such phase-coupled structure is constructed by two metallic stripes and an asymmetric waveguide with gain-assisted dielectric core and metallic substrate. We find that the lasing condition in the plane of the gain coefficient and the detuning is well-defined by a quadratic function. This quantitative function can be predicted by both simulation and coupled mode theory. We also find that in amplification regime the group delay and the transmittance of the SPP wave increase simultaneously as the detuning goes to zero. Our findings could be used to control the group velocity of SPP wave in the plasmonic waveguide in the field of signal processing and optical communications.

**Acknowledgement**

This work is supported by National Natural Science Foundation of China (11274396, 11074311, 61235002, 10804131), the Fundamental Research Funds for the Central Universities (12lgzd04), the Guangdong Natural Science Foundation (S2012010010537) and City University of Hong Kong (9360128 and 7008079).

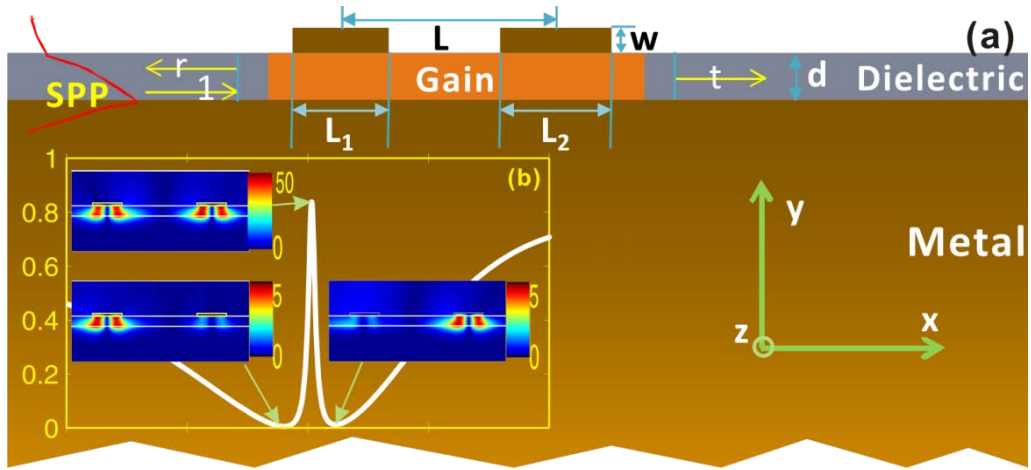

FIG. 1. (Color online) (a) Schematic of the sub-wavelength structure with two side-coupled resonators. The structure is constructed by two metallic stripes and an asymmetric waveguide with gain-assisted (orange region) dielectric core and metallic substrate. The stripes have same thickness (w) and slightly different lengths ($L_1$, and $L_2$) to realize frequency detuning. Their center-to-center spacing is L, and the thickness of the dielectric core is d. The SPP wave (red curves) propagates along the x direction with transverse magnetic polarization. (b) PIT transmission spectrum and the corresponding magnetic field ($|H_z|$) near two stripes at the peak and dips.

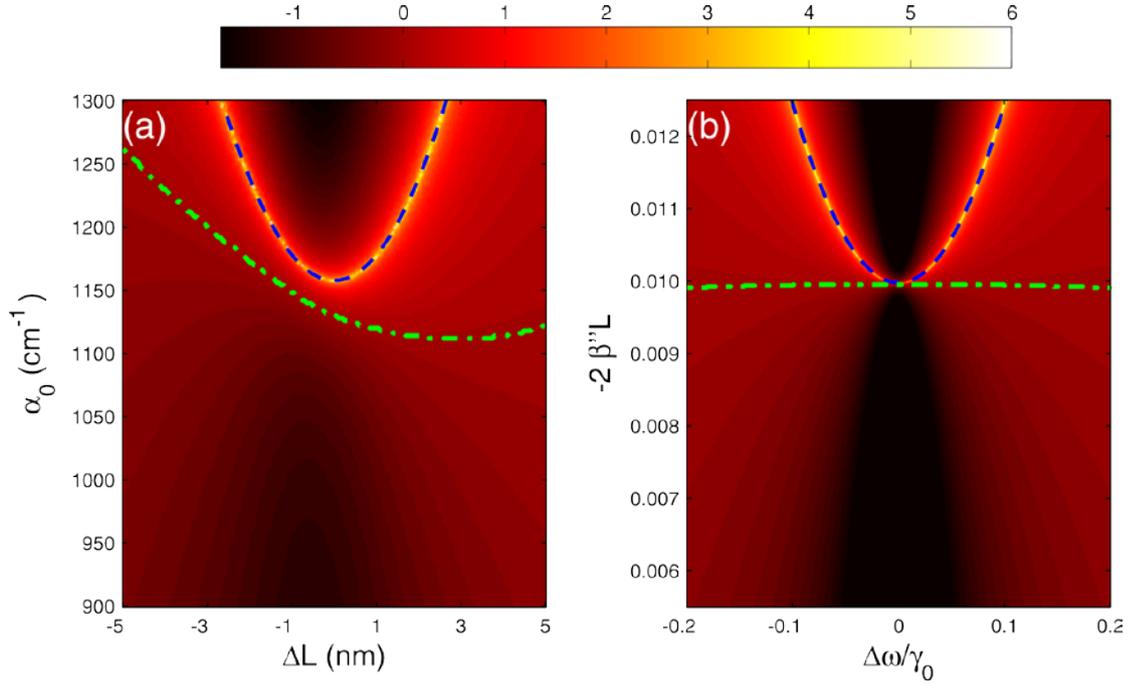

FIG. 2. (Color online) (a) Phase map in the plane of the gain coefficient ($\alpha_0$) and the frequency detuning ($\Delta L$), calculated by COMSOL. (b) Phase map in the plane of $-2\beta''L$ and the normalized frequency detuning ($\Delta\omega/\gamma_0$), calculated by coupled model theory. Note that $\beta''<0$ represents gain. In (b), we set $\gamma_c=\gamma_0$, $\gamma=0.005\gamma_0$, where $\gamma_0$ is a normalization factor. Colors show the value of $\log_{10}(T_{max})$, where $T_{max}$ is the peak transmittance of the PIT spectra. Green dash-dot curves indicate the $T_{max}=1$ contour. Blue dashed curves indicate the lasing threshold when the transmission coefficient diverges, obeying same form of functions, Eqs. (1) and (5), respectively.

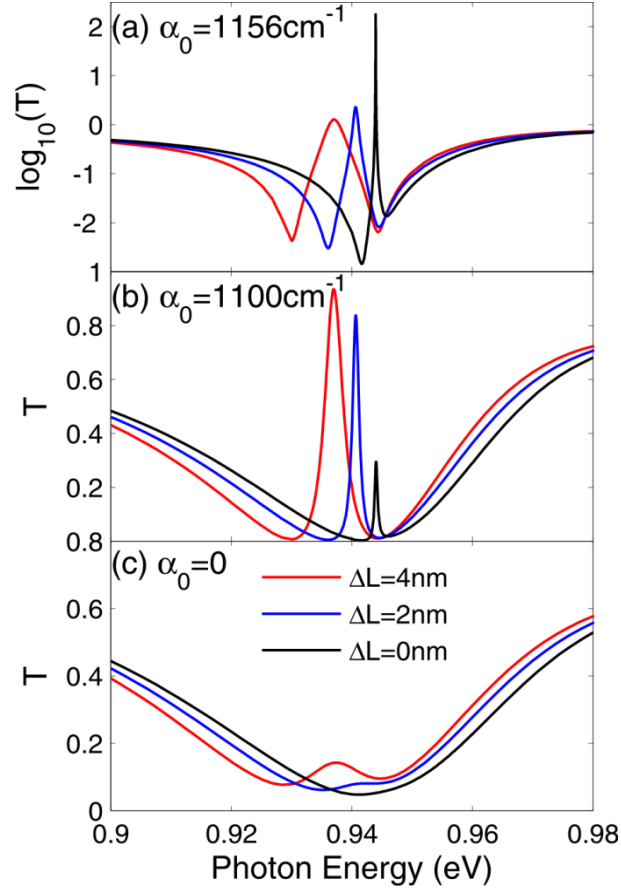

FIG. 3. (Color online) PIT transmission spectra (a) in amplification regime, (b) in loss-compensation regime, and (c) without gain. Color curves are for different detunings with the value of $\Delta L$ = 4 (red), 2 (blue), and 0 nm (black), respectively. Here, $L_2$ is fixed as 180 nm. Note that the vertical axis of (a) is log scale, indicating that the peak linewidth are much narrower than those in (b) and (c).

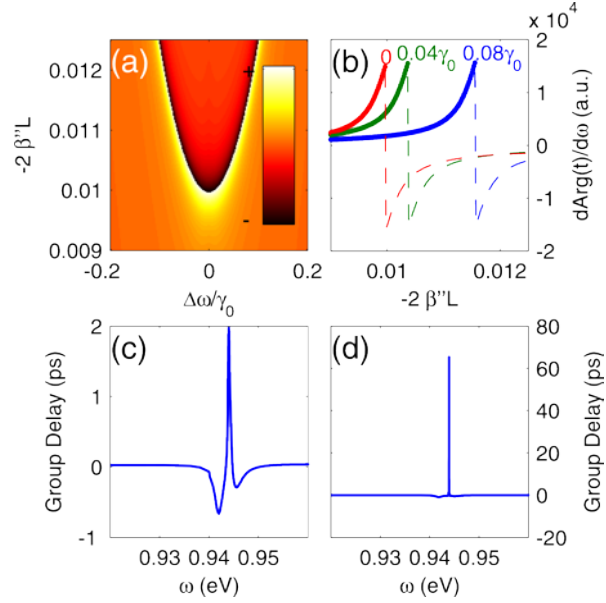

FIG. 4. (Color online) (a) Phase map in the plane of *-2β″L* and the detuning of two resonators ($\Delta\omega/\gamma_0$), calculated by coupled model theory. Here, we set $\gamma_c=\gamma_0$, $\gamma=0.005\gamma_0$, $\Phi_{rt}=0.8(\omega-\omega_0)+2\pi$, where, $\gamma_0$ is a normalization factor. Color in (a) shows the value of dArg(t)/dω, where Arg(.) is the argument of transmission coefficient, and d(.)/dω is derivative at the peak transmittance point in the spectra. (b) Vertical cuts of (a) at $\Delta\omega/\gamma_0=0$ (red), 0.04 (green), 0.08 (blue), respectively. (c) and (d) are group delay vs frequency of the structure in FIG. 1 with zero detuning and with gain coefficient of 1100 cm$^{-1}$ and 1156 cm$^{-1}$ respectively.